# Supporting Students Improve with Rubric-Based Self-Assessment and Oral Feedback

Sebastian Barney, Mahvish Khurum, Kai Petersen, Michael Unterkalmsteiner, Ronald Jabangwe

*Abstract*—Rubrics and oral feedback are approaches to help students improve towards learning outcomes. However, their effect on the actual improvement achieved is inconclusive. This paper evaluates the effect of rubrics and oral feedback on student's learning outcomes. An experiment has been conducted in a software engineering course on requirements engineering, using the two approaches on assignments in the course. Both approaches led to statistically significant improvements, though no material improvement was achieved (i.e. a change by more than one grade). The rubrics led to a significant decrease in the number of complaints/questions regarding grades.

## I. INTRODUCTION

THIS paper is motivated by the authors' observation that students feel that they deserve a higher grade, and hence ask for explanations. One potential reason is the gap in perception between teachers and students of what is expected in an assignment or an exam. When closing this gap a reasonable assumption is that students improve with respect to learning outcomes as they know what is expected.

Two well known methods are used as interventions to close the gap between the perception of teachers and students, and it is evaluated whether this improves the learning outcomes.

The first intervention is self-assessment using rubrics. Rubrics are split into several criteria (e.g. writing, selection of alternatives, reflection on solutions, and so forth). For each criteria levels are defined representing the level of understanding, preferably following the Structure of Observed Learning Outcomes (SOLO) taxonomy [1]. Research has shown that the impact of rubrics and self-assessment on learning outcomes is inconclusive [2], though it is acknowledged that rubrics have the potential of improving student's performance [3]. In addition, there are very few studies that are related to software engineering or computer science, which is the focus of the course in this study. Hence, there is a need to investigate and contribute further evidence by evaluating rubric based evaluation and self-assessment in software engineering courses.

The second intervention is verbal feedback. Studies report that students desire verbal feedback and perceive it as useful in order to improve [4]. In fact, they desire to receive feedback before assignment submissions [5]. The actual effect on learning outcomes and grades, however, is an open research question (cf. [5]). Hence, in this study the intervention is used as desired by the students in the previously mentioned studies, i.e. before submitting the final version of an assignment. This study makes a contribution to the research gap by evaluating the improvements achieved by the students.

The interventions are applied in a master's level course on large-scale requirements engineering in Sweden, with a total 67 participants. In Sweden students are allowed multiple submissions of each assessment task, and there are limitations imposed on the amount of time given to teachers to grade each assessment task. The two interventions are: 1) rubrics-based self evaluation, 2) teacher's feedback.

The remainder of the paper is structured as follows. Background and related work is introduced in Section II. The research design is presented in Section III, with results detailed in Section IV. A discussion of the results is made in Section V, and conclusions are drawn in VI.

## II. BACKGROUND AND RELATED WORK

This section presents a literature review regarding the two interventions to be empirically evaluated in this study. The first intervention is the introduction of rubrics for self-evaluation to improve student's learning. The second intervention is oral feedback on student's work. The review focuses on rubrics and feedback in general, not only for the higher education. One might argue that the difference between older and younger students is that the older students would like to have more independence in how they learn, i.e. they are likely to be more self-directed. However, as pointed out in [6] there are adults that are highly dependent on structure and guidance, while at the same time there are children that are self-directed learners. Hence, rubrics are potentially useful for all levels of education, and empirical results are (partially) generalizable between them. This is also indicated by reviews on the topic as they are including primary, secondary, and tertiary education in their assessment (see e.g. [7], [2]).

### A. Rubrics for Self-Evaluation

Rubrics represent the cognitive level achieved by a student with respect to learning outcome (see Figure 1 for the rubric used in this study). Rubrics should be designed according to the SOLO taxonomy where the level of achievement in the rubric maps to the levels in the Structure of Observed Learning Outcomes (SOLO) taxonomy [1]. The levels of the SOLO taxonomy range from pre-structural (presentation and reporting of incoherent information) to extended abstract (connecting information within a topic area and being able to

Sebastian Barney, Mahvish Khurum, Kai Petersen, Michael Unterkalmsteiner and Ronald Jabangwe are with Blekinge Institute of Technology, Karlskrona, Sweden. E-mails firstname.lastname@bth.se

Kai Petersen is also affiliated with Ericsson AB, Box 518, SE-371 23, Karlskrona, Sweden

transfer/generalize the information to other areas). In between those levels, there are different degrees of connection between information.

First, the authors focus the review on self-evaluation in general without a particular focus on rubrics. Ross [7] reviewed literature regarding the reliability of self-assessment and its usefulness in improving student performance. In his review Ross [7] considered primary, secondary, and tertiary education (i.e. from middle school to university level). With regard to reliability the literature review evaluated whether students are consistent in their self-evaluation across different tasks (intra-rater reliability). Overall, the findings indicate that students are consistent. Particularly students who have been trained in evaluation criteria were consistent. The time between tasks and the age of the students being evaluated impacts consistency, older students (14-16 years) are more consistent than younger students. When comparing self-assessment results with teachers' assessment or peer assessment (inter-rater reliability) the results vary. The results showing high inter-rater reliability can be criticized regarding their validity (unclear evaluation criteria, few replications, groups of students not comparable). The review of literature also revealed that students have a tendency to overestimate themselves, have interest bias, or are unable to apply assessment criteria. Hence, a factor that can increase the inter-rater reliability is when students are trained in self-assessment. With regard to improvements with regard to learning outcomes two groups of studies are identified. One group reports positive results arguing that self-assessment positively affects self-efficacy, motivation, and hence leads to stronger achievement. The other group reports negative results as students might select unrealistic goals, and hence adopt ineffective learning behavior. Or, if the goals are clearly given and they feel they cannot achieve them, this can lead to loss in motivation, and hence grades would not improve.

The review by Jonsson and Svingby [2] is very similar to the review of Ross [7], both aim at answering the same research questions regarding inter-rater relability and improvement. However, in [2] the focus is placed on rubric-based evaluation and self as well as peer assessment. The review also includes literature from middle school to university level. With regard to inter-rater reliability the review found that students using rubrics are reasonably consistent with values above 70%. With regard to agreement the results are more reliable than without rubrics. However, as pointed out by Jonsson and Svingby this does not necessarily make the evaluation better, as the rubric might only align views. At the same time the rubric might be flawed with respect to reflecting the desired learning outcomes in a good way. The results related to students improving through rubrics is inconclusive. Some studies report on an overall improvement. One example of such a study is [3] showing improvements in subsequent assignments within a course, as well as in comparison to results from previous years. The study by Green and Bowser [8] included in the review is of particular interest as it focuses on higher education and students conducting literature review and analysis, which is very similar to the task the students in this study have to conduct. Green and Bowser's study showed that students improve on some rubric criteria (three improvements), while not showing any change on two criteria, and even a negative impact on five criteria. The comparison was made by sampling reviews from two groups, one guided by rubrics and one not guided by rubrics.

Andrade [9] provided an experience report of the usage of rubrics. Based on her experience a number of conclusions with respect to rubrics have been drawn. Rubrics make it easier for the teacher to explain what is expected, and students feel that rubrics are useful. This was evident as students asked for rubrics as soon as they were used to them. Rubrics also help in explaining expectations clearly, i.e. they are not hidden anymore. In fact, often teachers have the assumptions that the students should know what is "good" and what is "poor", however, this is often not the case. Splitting rubrics into categories also supports students in identifying their strengths and weaknesses. Overall, based on conversations with students the teacher experienced that students learned more content with the introduction of rubrics.

Looking at the computer science and software engineering literature only few studies on self-assessment and rubrics can be found (cf. [10]). For computer science a web application development project used rubrics defined together with students. The results were that rubrics (1) help in defining ones own achievement goals (i.e. desired outcome); (2) are good at assessing against self-defined standards; (3) lead to higher satisfaction with grades. Furthermore, rubrics have been used in grading essays on ethics in computer science [11]. To test the rubrics two teachers graded one assignment independently and compared the results. There were only slight differences in grading. In outcome, the rubrics improved the criterion "English Language" while another criterion "Technical Details" went down. The explanation provided was that students noticed that language is important. A reason for the negative change for the criterion "Technical Details" was explained by the nature of the first assignment. In the first assignment the students could choose their own topic for the essay. which was not the case in the second assignment.

In software engineering rubrics have been suggested as a tool in higher education (cf. [12], [13]), but no detailed evaluations and empirical results were reported.

### B. Feedback (Verbal/Oral)

Blair and McGinty [4] conducted a study using action research at two universities focusing on political science/history students. The study was motivated by the observation that students have problems to understand feedback, and teachers have problems to provide good explanations that help students understand. A large portion of students at the two studied universities are convinced that verbal feedback helps them improve their learning. In fact, students desired to receive feedback after exams, and also receive feedback prior to the submission of assignments. However, according to Blair and McGinty [4] there exists a gap between the desire of students to get oral feedback, and the willingness of staff to provide that. In that situation, the authors encourage the use of verbal peer feedback, e.g. through the discussion of examples of essays.

Gibbs et al. [14] surveyed two universities (A and B) focusing on the programs in physics, chemistry, and bio chemistry. Teachers at university A mainly provided written feedback, while university B focused on oral feedback. Their findings are that from an effort perspective oral feedback is better for the teacher as it requires less time in comparison to writing detailed feedback. However, a problem observed with oral feedback was that it is not long-lasting as students can not easily refer back to the information provided. Also, there is a difference in perception of what feedback is. Teachers felt that they provide feedback all the time (e.g. in lectures, laboratories, workshops, and informally). However, the students do not count that information as feedback. This might also be a reason why the students do not undertake proper effort in documenting and storing the feedback provided.

Jollands et al. [5] investigated social sciences/engineering classes on university level by discussing in focus groups of 10 to 15 students. Their goal was to find out what good feedback is (written and verbal). For verbal feedback they defined criteria of when the feedback is successful. For example, in groups verbal feedback is only successful if it is focused on the knowledge gap of the student. In the class room the feedback is only useful if the answers encourage the students, and are constructive. Verbal feedback in class discussions is only useful if the students provide good and constructive comments leading to a valuable discussion. However, verbal feedback might lead to neglecting students that are less likely to approach the teacher, are not confident to ask questions, are not attending the class, and so forth. An open question is if good feedback in the class will increase grades, which was proposed as a future direction for research.

## III. Research Design

### A. Context (Course and Students)

The experiment was conducted in an academic setting, with 42 engineering graduate students at Blekinge Institute of Technology. It was conducted as a mandatory although non-graded exercise during a 7.5 ECTS merits master's course in large-scale requirements engineering (LSRE). Participation was mandatory and despite the ethical issues of forcing subjects to participate in a study, it was believed that the interventions introduced had several pedagogical benefits in the course. The students were instead given the option to exclude their individual results from the study, an option not utilized by any student.

### B. Research Questions

The main aim of this paper is to determine if rubric-based self-assessment and teacher's verbal feedback can be used to improve student learning outcomes. The first research question and set of hypotheses address rubric-based self assessment.

> **RQ1:** Can rubric-based self-assessments be used to support students improve with respect to learning outcomes?

The hypotheses tested in answering this question is thus:

$H_{10}$: The use of rubric-based self assessment does not change students' ability with respect to learning outcomes.

$H_{1a}$: The use of rubric-based self assessment can change students' ability with respect to learning outcomes.

The second research question and set of hypotheses address teacher's verbal feedback.

> **RQ2:** Can teacher's verbal feedback be used to support students improve with respect to learning outcomes?

The hypotheses tested in answering this question is thus:

$H_{20}$: The use of teacher's verbal feedback does not change students' ability with respect to learning outcomes.

$H_{2a}$: The use of teacher's verbal feedback can change students' ability with respect to learning outcomes.

### C. Design and Instrumentation

There are two assignments in the LSRE course. Each was broken into two parts, with the intervention applied in between. All students were subject to the same treatments for each assignment.

- Assignment 1 (A1) aimed to get students to reproduce the concepts learned. Students are not required to reflect or perform critical analysis in A1. A1 was to be done individually.
- Assignment 2 (A2) required students to reflect and critically analyze solutions they propose. A2 is done in pairs.

A1–Part 1: Description of A1 was given to the students. However, the students were not made aware of the intervention. The evaluation rubric for the assignment was made available on the course homepage and is shown in Figure 1. The students were given a deadline to submit A1–Part 1.

A1–Part 2: Shortly after the deadline for A1–Part 1, the students were informed about A1–Part 2 by the teacher. For this part students needed to assess their own assignment against the rubric and update their assignment based on this experience. Students were given two weeks to submit the rubric-based assessment and updated assignment. Students were also made aware that their course grade would only be based on the assignment submitted as A1–Part 2

A2–Part 1: A description of A2 was given to the students, however, the students were not made aware of the intervention. The evaluation rubric was also made available on the course homepage. The students were told to present their work in a 30-minute session, during whichthe teacher would provide live feedback on their assignments with an emphasis on the reflections and critical analysis parts of the assignment. Students were given a deadline for A2–Part 1, and assigned to presentation times one day after submission of this part.

A2–Part 2: After the presentation and teacher's feedback each group was told to use this feedback to update their initial submission. The students were given a deadline for this task, and informed that their final grade for this task would be based entirely on A2–Part 2

| | A | B | C | D | E |
|---|---|---|---|---|---|
| Context (MDRE and bespoke) | When, Why, Characterize, Explain, Properly referenced | When XOR why, Characterize, Explain, Properly referenced | When XOR why, Shallow explain, Properly referenced | When XOR why, Explain, Not properly referenced | Unclear, Not properly explained, Not properly referenced |
| Differences between MDRE and bespoke | Many differences, Describe, Characterize, Exemplify, Properly referenced | Many differences, Describe, Characterize, Properly referenced | Many differences, Shallow describe, Properly referenced | Few differences, Describe, Properly referenced | Only list some, Unclear, Not properly explained, Not properly referenced |
| Challenges | Many challenges, Describe, Characterize, Exemplify, Properly referenced | Many challenges, Describe, Characterize, Properly referenced | Many challenges, Shallow describe, Properly referenced | Few challenges, Describe, Properly referenced | Only list some, Unclear, Not properly explained, Not properly referenced |
| Analysis and Discussion | Conclusions (deep), Based on references, Own experiences, Separate section, Properly referenced | Conclusions (deep), Based on references, Separate section, Properly referenced | Conclusions (not deep), Based on references, Separate section, Properly referenced | Conclusions, Based on references, Separate section, Not properly referenced | Unclear, Not properly explained, Not properly referenced |
| English | No serious, Less than 2 minor errors | No serious errors, Few minor errors (less than 5) | Occasional serious errors, Many minor errors | Frequent errors, Many serious errors, But still readable | Not readable, Multiple, serious errors |
| References | More than 10 peer rev., Primary, Relevant | More than 10 peer rev., Primary, Some not really relevant | Minimum 5 peer rev., Some non peer rev. | Minimum 5 peer rev., Some are not referred to | Less than 5 peer rev. |
| Contextual flow in the assignment | Excellent flow with separate sections for introduction, context description, challenges, analysis and discussion | Nice flow with separate sections for introduction, context description, challenges, analysis and discussion | Nice flow, some sections not separated/missing | Some flow, some sections not separated/missing | No flow, a lot of sections missing |

Fig. 1. Rubric for Assignment 1

## D. Data Collection and Analysis

Data was collected at a number of points during this course as part of this study:

1) After the introductory lecture, a questionnaire was used to collect students' desired and expected grades for each assessment task in the course.
2) A1–Part 1 was graded by a teacher, however, at no stage was this grade made known to the students.
3) For A1, the students' grades from their rubric-based self-assessments were collected.
4) A1–Part 2 was graded by the same teacher responsible for grading A1–Part 1.
5) A2–Part 1 was graded by a teacher, however, at no stage was this grade made known to the students.
6) Identifiers for students who presented their work and received verbal feedback from the teacher were recorded.
7) A2–Part 2 was graded by the same teacher responsible for grading A2–Part 1.
8) A post-course questionnaire asked students to rate the helpfulness of the two interventions in improving their grade.

To test if a statistically significant improvement was made between Part 1 and Part 2 in each assignment, two-tailed paired t-tests were applied to the grades in each group.

## IV. RESULTS

### A. Intervention 1: Rubric-Based Self-Assessment

The results of this section are limited to the 40 students that completed Part 1, the rubric-based self reflection and Part 2 for Assignment 1.

At the start of the course students were asked about the grades they desired and expected to achieve for each assessment task in the course. The results for Assessment 1 are shown in Figure 2, with almost all students desiring and expecting an *A*, *B* or *C*. The students' expected grade distribution is slightly lower than their desired grade distribution.

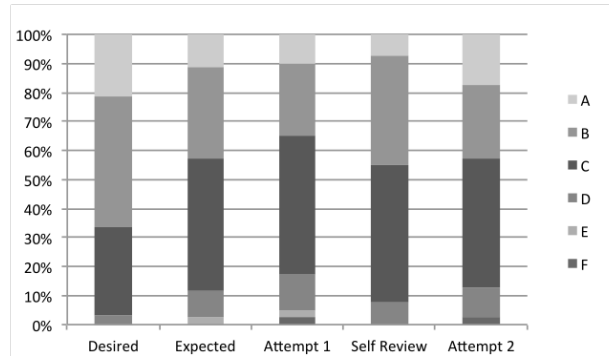

Fig. 2. Grade distributions for Assignment 1

The distribution of grades for Part 1, the Self Reflection and Part 2 for Assignment 1 are shown in Figure 2. The distribution of grades for Part 1 is slightly below the students' expected grades. At no stage where students made aware of their grade for Part 1.

The students were effective at assessing their own work against the rubric for the Self Reflection exercise, awarding grades *A* through *F*. The students self-assessment matched the teachers grade in 45% of cases and was accurate to within one level for 85% of cases. The remaining 15% of students differed by two grade levels. Students were equally likely to overestimate and underestimate their grades.

Students were given two weeks to complete the Self Reflection exercise and submit Part 2 for Assignment 1. The results saw 9 of the 40 students improve their grade one level between Part 1 and Part 2 (eg. $B \rightarrow A$), with the remainder of the students seeing no material change in their grades.

This result showed a statistically significant change in the grades using a two-tailed paired *t*-test ($p = 0.0017$), allowing the null hypothesis to be rejected ($H_{10}$). However, this reflection on their own work did not translate into a large improvement in terms of the grades assigned, as shown in Table I.

TABLE I
GRADE DISTRIBUTIONS AND AVERAGE GRADES

| Grades | A1-P1 | A1-P2 | A2-P1 | A2-P2 |
|---|---|---|---|---|
| A (95.5) | 4 | 7 | 0 | 2 |
| B (85.5) | 10 | 10 | 2 | 0 |
| C (75.5) | 19 | 18 | 2 | 4 |
| D (65.5) | 5 | 4 | 0 | 0 |
| E (55.0) | 1 | 0 | 5 | 7 |
| F (24.5) | 1 | 1 | 10 | 6 |
| *Average* | 77 | 79 | 44 | 53 |

Students perceived the rubric-based intervention as very helpful for improving their grade. In a post course questionnaire students were asked to what degree the agreed with the statement that the rubric-based intervention was helpful in improving the assignment. Responses were taken in the form of a five point Likert scale (1=strongly disagree, 5=strongly agree), with an average response of 4.7. The questionnaire was completed by 9 of the 40 students who completed Part 1, the Rubric-Based Self Assessment and Part 2 of Assignment 1.

### B. Intervention 2: Teacher's verbal feedback

The results of this section are limited to the 19 students that completed Part 1, the Presentation and Part 2 of Assignment 2.

Students were asked about the grades they desired and expected to achieve for each assessment task at the start of the course. The results for Assessment 2 are shown in Figure 3, with all students desiring and expecting an *A*, *B* or *C*. The distribution students' desired and expected grades is higher than with the first Assignment, despite this being a larger and more complex assignment. Further, the expected results are higher than the desired results, indicating the students expect to exceed their desires in terms of the grade on this assignment.

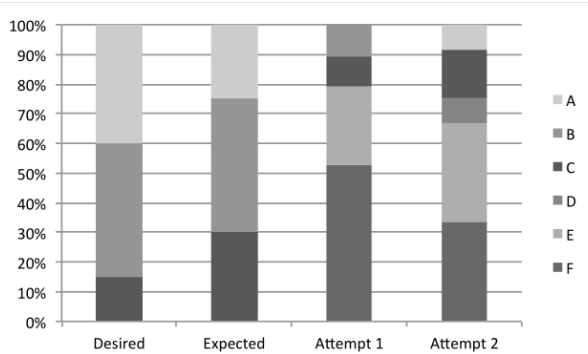

Fig. 3. Grade distributions as part of Assignment 2

The distribution of grades for Part 1 and Part 2 of Assignment 2 are shown in Figure 3. The distribution of grades for Part 1 is far below the students' desired and expected grades with 10 out of 19 students failing. The highest grade awarded two *B*s, followed by two *C*s and five *E*s. At no stage where students made aware of their grade for Part 1.

After completing Part 1, students made an oral presentation to the lecturer to receive feedback on their assignment. After receiving this feedback students were given two weeks to update their assignments.

Part 2 saw a statistically significant change from Part 1 using a two-tailed paired t-test ($p = 0.0044$)—allowing the null hypothesis to be rejected ($H_{20}$). As summarized in Table I, two students raised their grades two levels (ie. $E \to C$), and six students raised their grade one level (ie. $2 \times B \to A$ and $4 \times F \to E$). The remaining 11 students saw no material change in their grade.

With this assignment it was also possible to determine the amount of changed material between Part 1 and Part 2 of the Assignment. The students who improved their *B* to an *A* did so by only changing 3% of their assignment, the smallest change seen between the two parts. The students who saw a material improvement in their grade changed on average 63% of their assignments for the Part 2, while the students who did not see any material improvement in their grade changed on average 54% of their assignments for Part 2.

Of the 19 students who completed Assignment 2 as planned, 10 respondent to the post course questionnaire. The students strongly agreed with the statement that "the presentation exercise and feedback from the teacher as part of Assignment 2 was very helpful in approving the assignment." Students answered on a five-point Likert scale (1=strongly disagree, 5=strongly agree), with an average result of 4.5.

### C. Grading Complaints

A significant change was also seen in the number of complaints about grades. In this implementation of the course there was only one complaint about grades. Previous years, that have not included these interventions to help students improve their grades, have seen approximately 10 complaints about grades.

## V. DISCUSSION

As shown in the survey, from the beginning of the course, students have high desires and expectations in terms of their grades. However, the results of the assignments show that the students are neither meeting their desires nor expectations. The fact that students expect to receive grades higher than they desire for the more complex A2, suggests they perceive that the course will be easy.

For students whose grades improved, it is possible that some of the improvement can be contributed to the additional time spent by students on the assignments as part of the interventions. However, for students who did not achieve any improvement in their grades, other factors might have reduced students' ability to spend the amount time on their assignment they perceived necessary, with students citing other courses, their masters' theses and personal reasons as factors limiting their ability to take full opportunity of the interventions. Thus, as future work, authors are planning to conduct an exploratory case study to interview students from both samples (improved grades sample and no improvement sample) to investigate which factors related to the two interventions and which factors besides the interventions affected/not affected their grades in resubmissions. Further, it is important to emphasize that such interventions need to be planned, coordinated and applied across the masters program to reap the greatest benefits.

Further, it is possible that the students only improved in terms of performance against the rubric, and not in terms of the intended learning outcomes [2]. Some aspects assessed by the rubric are far more important than others at measuring students learning outcomes. For example, it is possible for a student to improve their grade through *English* while keeping a poor *analysis and discussion*; while another student who makes a significant improvements to the *analysis and discussion* may not see any material improvement in grade due to poor *English*. Thus, it needs to be further explored what aspects students improved in their assignments that led to an improvement in their grades.

Further, the success of each approach is dependent on the quality of the rubric and the teacher's oral feedback. Changes to either could increase or decrease students' ability to meet learning outcomes. From teachers perspective, the current system for assigning and controlling teaching hours does not support teachers to spend additional time helping students improve (for example, by spending more time to give oral feedback). A teacher can do so, but he/she will not receive additional teaching hours for this activity. Given such a constraint, the improvements in grades through the interventions become significant. This fact points to the need for the system to allow and accommodate efforts to better support students learning.

A significant result is the reduction in the number of complaints about grades. This result suggests that the interventions provided students with a greater understanding of what was expected from them from the assessment tasks. However, the teachers hoped this understanding would lead to a greater increasing in learning outcomes as measured through grades.

## VI. CONCLUSION

This paper evaluates two approaches to support students improve against course learning outcomes—rubric-based self assessment and teach's oral feedback. Both techniques are shown to lead to improved learning outcomes when applied to a completed assignment and students are given time to address the feedback provided. The increase in learning outcomes, however, was much more limited than the teachers expected.

The major change seen from the implementation of these interventions was an increase in student understanding of teachers' expectations. Compared to previous years, there were far fewer complaints about grades.

The rubric-based self assessment requires and investment of students' time, while the teacher's oral feedback requires an investment of both the students' and teachers' time. Given the improvement against learning outcomes was more limited than expected, it remains an open question as to whether the cost-benefit of these approaches is sufficient to justify their use. The author's recommend further empirical studies, preferably designed to allow comparisons between different approaches.

**Sebastian Barney** received his PhD in Software Engineering from Blekinge Institute of Technology.

**Mahvish Khurum** is a PhD student in Software Engineering at Blekinge Institute of Technology.

**Kai Petersen** is an industrial Post Doctoral researcher at Ericsson AB and Blekinge Institute of Technology.

**Michael Unterkalmsteiner** is a PhD student in Software Engineering at Blekinge Institute of Technology.

**Ronald Jabangwe** is a PhD student in Software Engineering at Blekinge Institute of Technology.